\documentclass[prl,showpacs,twocolumn,amsmath,amssymb]{revtex4}
\usepackage{epsfig}
\usepackage{graphicx}% Include figure files
\usepackage{dcolumn}% Align table columns on decimal point
\usepackage{color}
\usepackage{bm}% bold math

\begin{document}

\title{Impurity induced spin-orbit coupling in graphene}
\author{A. H. Castro Neto$^1$ and F. Guinea$^2$ }
\affiliation{
$^1$ Department of Physics, Boston University, 590 Commonwealth Ave., Boston MA 02215, USA
\\
$^2$ Instituto de  Ciencia de Materiales de Madrid, CSIC,
 Cantoblanco E28049 Madrid, Spain}

\begin{abstract}
We study the effect of impurities in inducing spin-orbit coupling in graphene. We show that the sp$^3$ distortion induced by an impurity can lead to a large
increase in the spin-orbit coupling with a value comparable to the one 
found in diamond and other zinc-blende semiconductors. The spin-flip scattering produced by the impurity leads to spin scattering lengths of the order found in recent experiments. Our results indicate that the spin-orbit coupling can be controlled via the impurity coverage.
\end{abstract}


\maketitle

Since the discovery of graphene in 2004 \cite{Netal04t} much has 
been written about its extraordinary charge transport properties
\cite{geim_review,NGPNG08}, such as sub-micron electron mean-free paths,
that derive from the specificity of the carbon $\sigma$-bonds against
atomic substitution by extrinsic atoms. However, being an open surface, 
it is relatively easy to hybridize the graphene's $p_z$ orbitals with
impurities with direct consequences in its transport properties
\cite{CJFWI08,blake08}. This capability for hybridization with external 
atoms, such as hydrogen (the so-called graphane), has been shown to 
be controllable and reversible 
\cite{graphane} leading to new doors to control graphene's properties.

Much less has been said about the spin-related transport properties such 
as spin relaxation, although recent experiments show that the spin 
diffusion length scales \cite{HJPJW07t,TTVJJvW08t} are much shorter than 
what one would expect from standard spin-orbit (SO) scattering mechanisms 
in a sp$^2$ bonded system \cite{HGB08}. In fact, atomic SO 
coupling in flat graphene is a very weak
second order process since it affects the $\pi$ orbitals only through
virtual transitions into the deep $\sigma$ bands \cite{HGB06}. 
Nevertheless, it would be very interesting if one could 
enhance SO interactions because of the prediction of the quantum spin Hall 
effect in the honeycomb lattice \cite{KM05} and its relation to the 
field of topological insulators \cite{KM07}. 

In this paper we argue that impurities (adatoms), 
such as hydrogen, can lead to a 
strong enhancement of the SO coupling due to the lattice distortions 
that they induce. In fact, it is well known that atoms that hybridize 
directly with a carbon atom induce a distortion of the graphene lattice 
from sp$^2$ to sp$^3$ \cite{DSL07}. By doing that, the electronic energy 
is lowered and the path way to chemical reaction is enhanced. Nevertheless, 
it has been known for quite sometime \cite{cardona} that in diamond, 
a purely sp$^3$ carbon bonded system, spin orbit coupling plays an 
important role in the band structure since it is a first order 
effect, of the order of the atomic SO interaction, 
$\Delta_{so}^{at} \approx 10$ meV, in carbon \cite{SCR00}. Here we show
that the impurity induced sp$^3$ distortion of the flat graphene lattice
lead to a significant enhancement of the SO coupling, explaining
recent experiments \cite{HJPJW07t,TTVJJvW08t} in terms of the      
Elliot-Yafet mechanism for spin relaxation \cite{Elliot54,Yafet63}
due to presence of unavoidable environmental impurities in the experiment.
Moreover, our predictions can be checked in a controllable way in graphane
\cite{graphane} by the control of the hydrogen coverage.

\begin{figure}
\includegraphics[width=8cm]{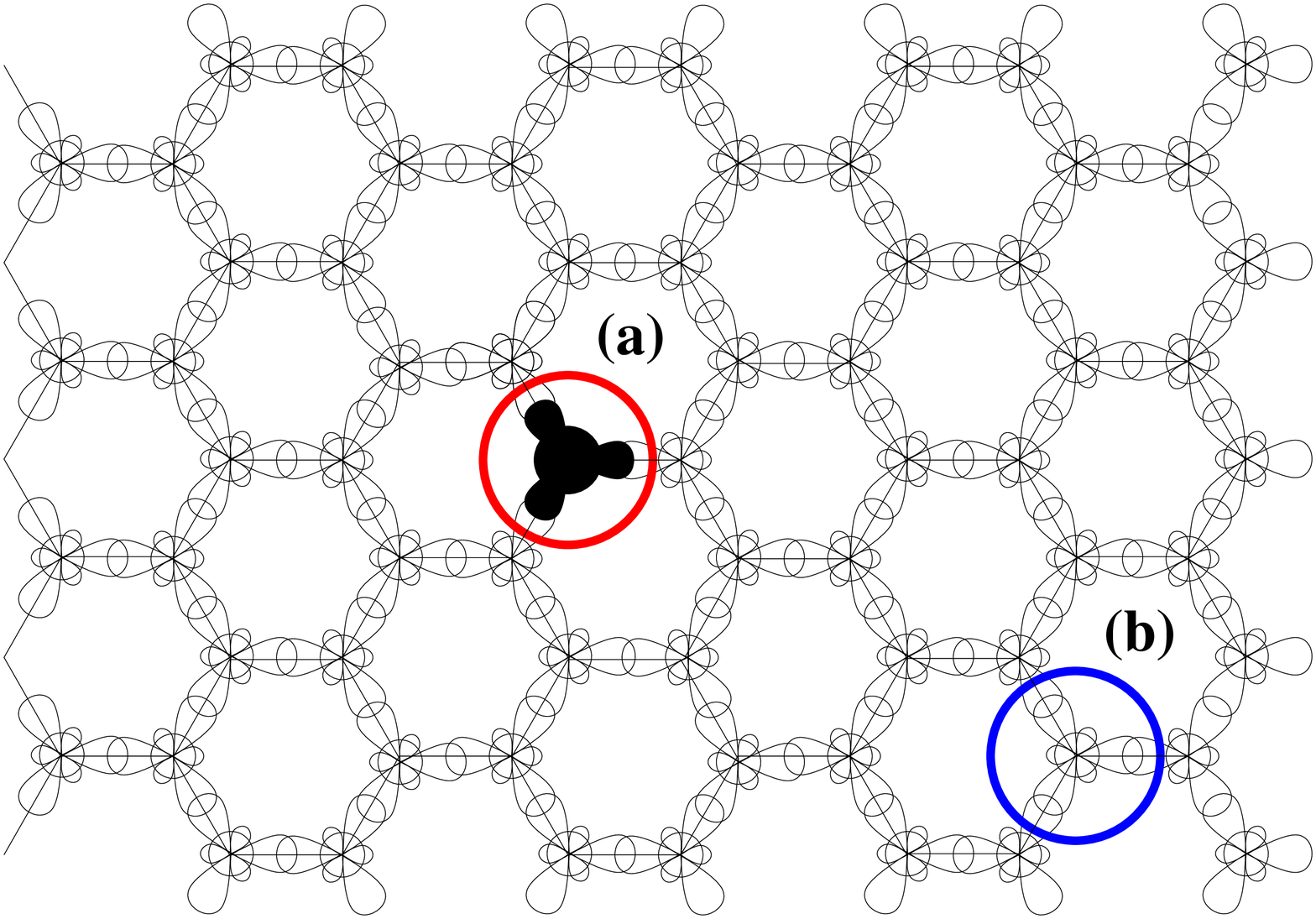}
\includegraphics[width=3.5cm]{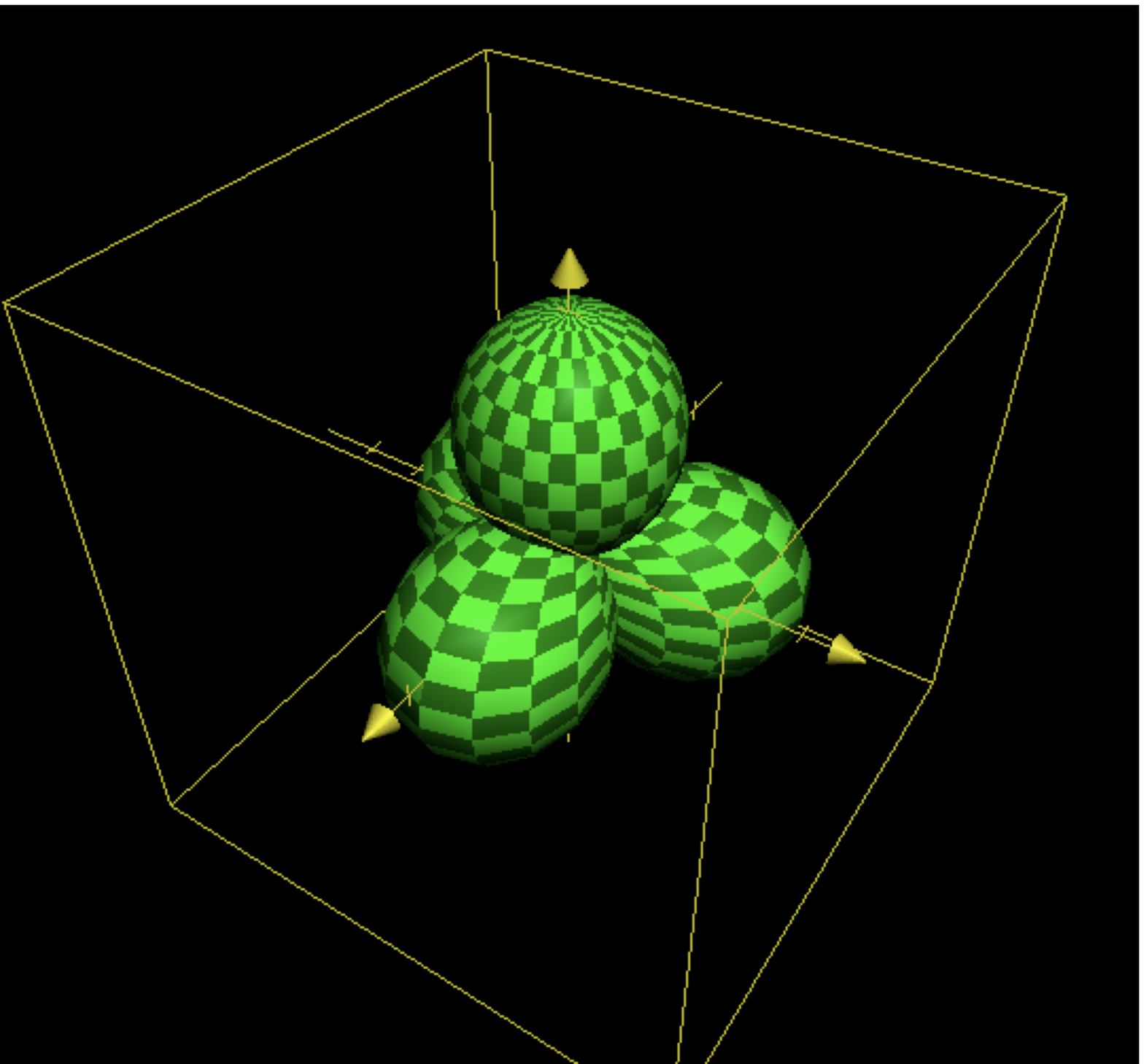} (a)
\includegraphics[width=3.5cm]{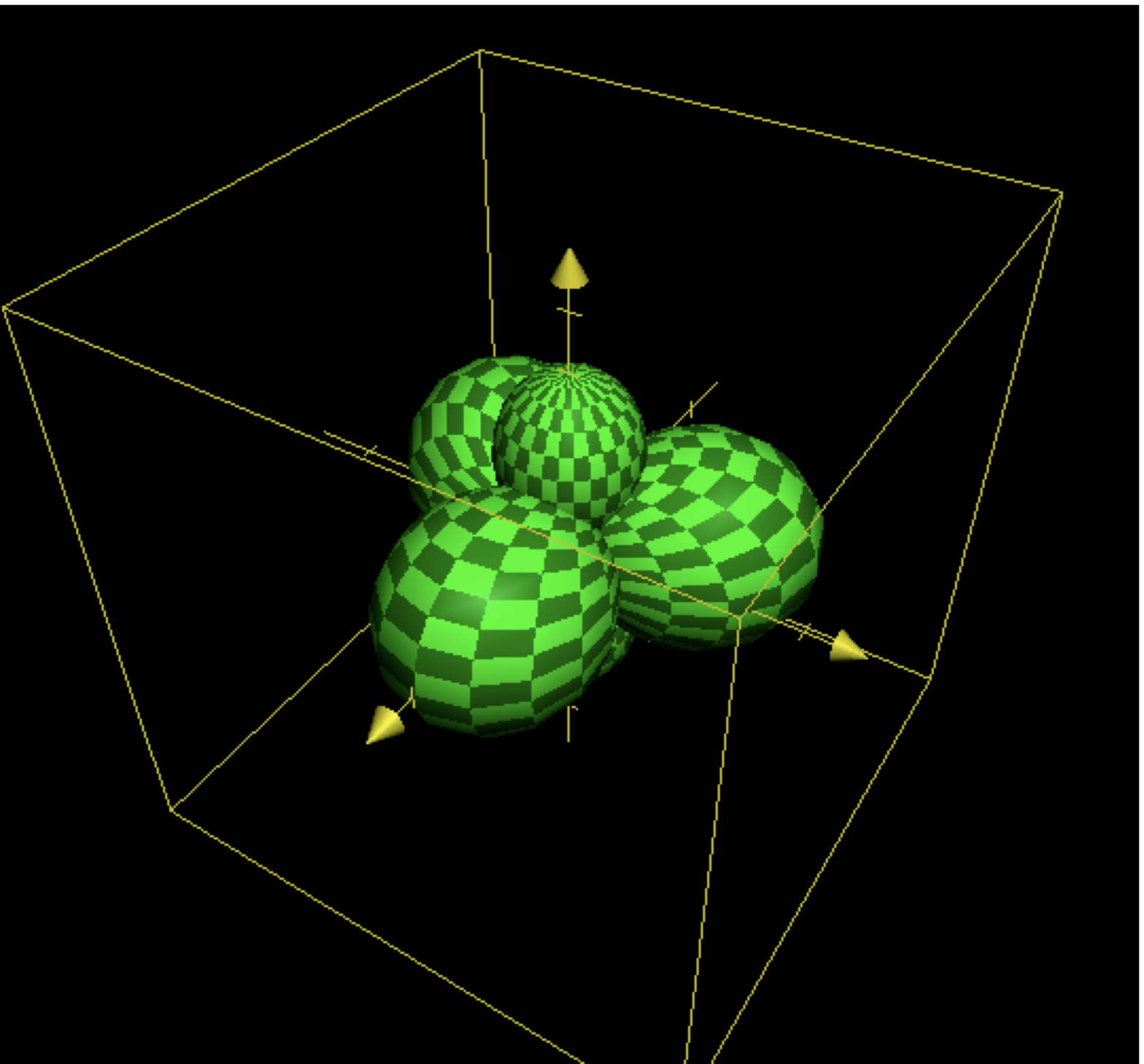} (b)
\caption[fig]{(Color online). Top: Top view of the 
graphene lattice with its orbitals. The orbitals associated
with the impurity and lattice distortion are shown in solid black. 
(a) sp$^3$ orbital at impurity position; 
(b) sp$^2$ orbital of the flat graphene lattice.} 
\label{structure}
\end{figure}

\begin{figure}
\includegraphics[width=8cm,angle=0]{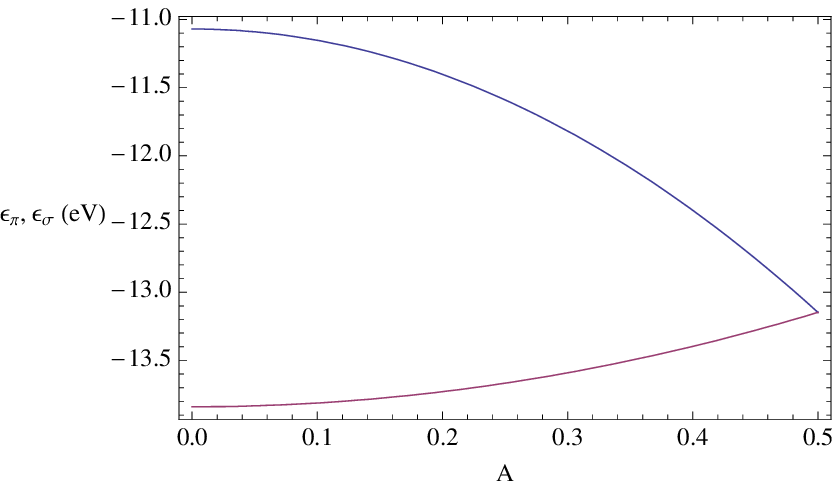}
\includegraphics[width=8cm,angle=0]{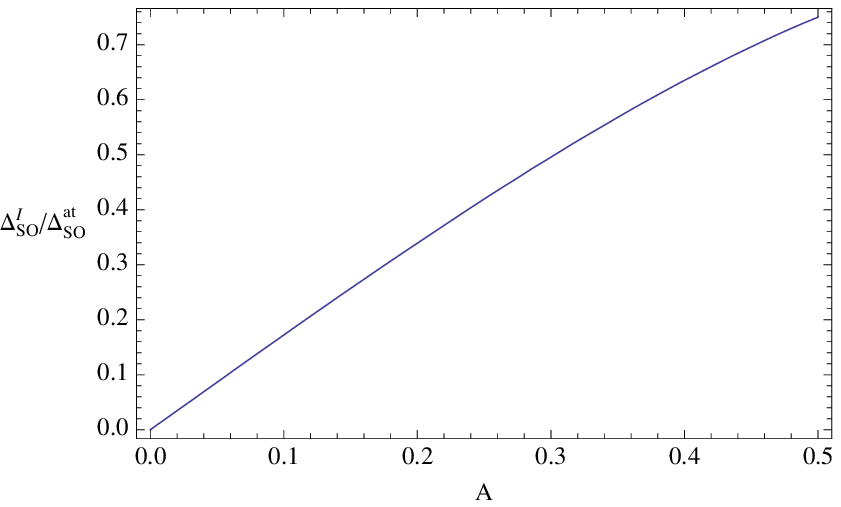}
\caption[fig]{(Color online).Top: Energy (in eV) of the $\pi$ (blue) and $\sigma$ (red) bands as a function of $A$ according to (\ref{epiesig}); Bottom: Relative value of the SO coupling at the impurity site relative to the atomic value in carbon as a function of $A$ according to (\ref{fim}).} 
\label{res}
\end{figure}

We assume that the carbon atom attached to an impurity is raised above 
the plane defined by its three carbon
neighbors (see Fig.~\ref{structure}). The local orbital basis at the position
of the impurity (which is assumed to be located at the origin, 
${\bf R}_{i=0}=0$) can be written as:
\begin{align}
| \pi_{i=0} \rangle &= A | s \rangle + \sqrt{1-A^2} | p_z \rangle \, ,
\nonumber \\
| \sigma_{1,i=0} \rangle &= \sqrt{\frac{1-A^2}{3}} | s \rangle -
\frac{A}{\sqrt{3}} | p_z \rangle + \sqrt{\frac{2}{3}} | p_x \rangle \, ,
\nonumber \\
| \sigma_{2,i=0} \rangle &= \sqrt{\frac{1-A^2}{3}} | s \rangle -
\frac{A}{\sqrt{3}} | p_z \rangle - \frac{1}{\sqrt{6}} | p_x \rangle
+ \frac{1}{\sqrt{2}} | p_y \rangle \, ,
\nonumber \\
| \sigma_{3,i=0} \rangle &= \sqrt{\frac{1-A^2}{3}} | s \rangle -
\frac{A}{\sqrt{3}} | p_z \rangle - \frac{1}{\sqrt{6}} | p_x \rangle
- \frac{1}{\sqrt{2}} | p_y \rangle \, ,
\label{basis}
\end{align}
where $|s\rangle$, and $|p_{x,y,z}\rangle$,  
are the local atomic orbitals. 
Notice that this choice of orbitals interpolates 
between the sp$^2$ configuration, $A=0$, to the sp$^3$ configuration, 
$A= 1/2$. The angle $\theta$ between the new $\sigma$ orbitals and 
the direction normal to the plane is $\cos(\theta) = -A/\sqrt{A^2+2}$.
The energy of the state $|\pi_i\rangle$, $\epsilon_{\pi}$, and
the energy of the three degenerate states $|\sigma_{a,i}\rangle$,
$\epsilon_{\sigma}$ ($a=1,2,3$), are given by (see Fig.~\ref{res}):
\begin{eqnarray} 
\epsilon_{\pi}(A) &=& A^2 \epsilon_s + (1-A^2) \epsilon_p \, ,
\\
\epsilon_{\sigma}(A)&=&(1-A^2) \epsilon_s/3+(2+A^2) \epsilon_p/3 \, ,
\label{epiesig}
\end{eqnarray}
where  $\epsilon_s \approx -19.38$ eV ($\epsilon_p \approx -11.07$ eV) is 
the energy of the $s$ ($p$) orbital \cite{harrison}. At the impurity
site one has $A \approx 1/2$ while away from the impurity $A=0$. 

The Hamiltonian of the problem can be written as,
${\cal H} = {\cal H}_{\pi} + {\cal H}_{\sigma}  + \delta {\cal H}$,
where ${\cal H}_{\pi}$ (${\cal H}_{\sigma}$) describes 
the $\pi$-band ($\sigma$-band) of flat graphene, 
and $\delta {\cal H}$ describes the {\it local} 
change in the hopping energies due to the presence of the impurity 
and sp$^3$ distortion:
\begin{eqnarray}
\delta {\cal H} &=& 
\sum_{\alpha=\uparrow,\downarrow} 
\left\{ \epsilon_{I} c^\dag_{I \alpha} c_{I \alpha} + t_{C-I} 
c^\dag_{I \alpha} c_{\pi \alpha 0}  
\right.
\nonumber
\\
&+&  \delta \epsilon_{\pi} c^\dag_{\pi \alpha 0} c_{\pi \alpha 0} + 
\delta \epsilon_{\sigma} \sum_{a=1,2,3} 
c^\dag_{\sigma_a \alpha 0} c_{\sigma_i \alpha 0} 
\nonumber 
\\
&+& \left. V_{\pi\sigma}
c^\dag_{\pi \alpha 0} \left( c_{\sigma_1 \alpha 0} + c_{\sigma_2 \alpha 0} +
c_{\sigma_3 \alpha 0} \right)+ {\rm h. c.} \right\} 
\label{hamil}
\end{eqnarray}
where 
\begin{eqnarray}
V_{\pi\sigma}(A)= A  \sqrt{\frac{1-A^2}{3}} (\epsilon_s - \epsilon_p ) \, ,
\end{eqnarray}
$c_{I,\alpha}$ ($c^\dag_{I,\alpha}$) annihilates (creates) an 
electron at the impurity, and $c_{\pi \alpha i}$
($c_{\sigma_a \alpha i}$) annihilates an electron at a carbon site 
in an orbital $\pi$ ($\sigma_a$) at
position ${\bf R}_i$ with spin $\alpha$, 
$\epsilon_I$ is the electron energy in the impurity,
and $t_{C-I}$ the tunneling energy between the carbon and impurity,
$\delta \epsilon_{\pi}(A) = \epsilon_{\pi}(A)-\epsilon_{\pi}(A=0)$,
and $\delta \epsilon_{\sigma}(A) = \epsilon_{\sigma}(A)-\epsilon_{\sigma}(A=0)$. 
In (\ref{hamil}) we have not included the change in the hopping between
$\sigma_{a,0}$ orbitals (the change in energy due to the distortion is 
$-A^2 (\epsilon_s-\epsilon_p)/3$) and the inter-atomic hopping terms. 
In this way, we have simplified the calculations and
the interpretation of the results. The inclusion of the other terms
do not modify our conclusions.

The atomic spin orbit coupling, 
${\cal H}_{so}^{at} = \Delta_{so}^{at} {\bf L} \cdot {\bf S}$, 
induces transitions between $p$ orbitals of different spin projection 
\cite{HGB06}. 
In flat graphene ($A=0$), it leads to transitions between the $\pi$
 and $\sigma$ bands. The change in the ground state 
energy in this case is rather small and given by:  
$(\Delta_{so}^{at})^2 / ( \epsilon_{\pi}(A=0) - \epsilon_{\sigma}(A=0)) 
\approx 10^{-2}$ meV \cite{HGB06}. 
However, the perturbation described by (\ref{hamil}) leads to a direct 
local hybridization $V_{\pi\sigma}$ between the $\pi$ and $\sigma$ bands 
that modifies the effective SO coupling acting on the $\pi$ electrons.
The propagator of $\pi$ electrons from position ${\bf R}_i$ with
spin $\alpha$ to ${\bf R}_j$ with spin $\beta$ can
be written as:
\begin{eqnarray}
\langle \pi_{i,\alpha} | \left( \epsilon - {\cal H} \right)^{-1}
| \pi_{j,\beta} \rangle \approx 
\langle \pi_{i,\alpha} |
\left( \epsilon - {\cal H}_\pi \right)^{-1} | \pi_{0,\alpha} \rangle
\nonumber 
\\  
\times \langle \pi_{0,\alpha} | \delta {\cal H}|\bar{\sigma}_{0,\alpha} \rangle 
\times 
\langle \bar{\sigma}_{0,\alpha}| \left( \epsilon - {\cal H}_\sigma \right)^{-1} |
\bar{\bar{\sigma}}_{k,\alpha} \rangle 
\nonumber \\ 
\times \langle \bar{\bar{\sigma}}_{k,\alpha} | {\cal H}_{so}^{at} 
|\pi_{k,\beta} \rangle 
\langle \pi_{k,\beta} | \left( \epsilon -
{\cal H}_\pi \right)^{-1} | \pi_{j,\beta} \rangle
\label{green}
\end{eqnarray}
where $|\bar{\sigma}_{0,\alpha} \rangle = [ | {\sigma_1}_{0,\alpha}
\rangle + | {\sigma_2}_{0,\alpha} \rangle + | {\sigma_3}_{0,\alpha}
\rangle ]/\sqrt{3}$ and $\bar{\bar{\sigma}}_{j,\alpha} \rangle  =
[ | {\sigma_1}_{j,\alpha} \rangle + e^{i \phi}|
{\sigma_2}_{j,\alpha} \rangle + e^{2 i \phi} | {\sigma_3}_{j,\alpha} 
\rangle ]/\sqrt{3}$ where $\phi = 2 \pi / 3$.
The propagator in (\ref{green}) can be understood as arising from
an effective non-local SO coupling within the $\pi$ band which goes
as:
\begin{eqnarray} 
\Delta_{so}^{I}(0,i) 
\approx V_{\pi\sigma} \langle \bar{\sigma}_{0,\alpha} | ( \epsilon -
{\cal H}_\sigma )^{-1} | \bar{\bar{\sigma}}_{i,\alpha} 
\rangle
\Delta_{so}^{at}  \, ,
\label{ind}
\end{eqnarray}
which allows us to estimate the local value of the SO coupling as: 
\begin{eqnarray}
\frac{\Delta_{so}^{I}(A)}{\Delta_{so}^{at}} 
%&\approx& A \sqrt{\frac{1\!-\!A^2}{3}} \!\left(\!\frac{\epsilon_s \!-\!
%\epsilon_p}{\epsilon_\sigma(A\!=\!0) \!-\! \epsilon_\pi(A\!=\!0)}\!\right)
%\nonumber
%\\
\approx  A \sqrt{3(1\!-\!A^2)}  \, .
\label{fim}
\end{eqnarray}
As shown in Fig.~\ref{res} the value of the SO coupling depends on the angle (i.e., the
value of $A$) associated with the distortion of the carbon atom away
from the graphene plane. Notice that for the sp$^2$ case ($A=0$)
this term vanishes indicating that SO only contributes in second
order in $\Delta_{so}^{at}$, while for the sp$^3$ case ($A=1/2$), 
the SO coupling is approximately $75 \%$ of the atomic value ($\approx 7$ meV). Also observe that the dependence on 
the distance from the location of the hydrogen
atom is determined by the Green's function $G_\sigma (0,{\bf R}_j)
) = \langle \bar{\sigma}_{0} | ( \epsilon - {\cal H}_\sigma
)^{-1} | \bar{\bar{\sigma}}_j \rangle$. This function,
calculated for the simplified model of the $\sigma$ bands discussed
in ref.~\cite{HGB06}, shows a significant dispersion in Fourier space,
ranging from a maximum at the $\Gamma$ point to zero at the $K$ and
$K'$ points. Hence, the range of $G_\sigma ( 0 ,{\bf R})$ should
be of the order of a few lattice constants. 

Based on the previous results we can now calculate the effect
of the impurity induced SO coupling in the transport properties.
Firstly, we linearize the $\pi$ band around the K and K' points
in the Brillouin zone and find the 2D Dirac spectrum \cite{NGPNG08}:
$\epsilon_{\pm,{\bf k}}= \pm v_F k$ where $v_F$ ($\approx 10^6$ m/s)
is the Fermi-Dirac velocity. In this long wavelength limit
the impurity potential induced by (\ref{ind}) has cylindrical symmetry
and we can use a decomposition of the wavefunction in terms of radial
harmonics \cite{OGM06,HG07,N07,KN07,G08}. A
similar analysis, for a system with SO interaction in the
bulk has been studied in ref.~\cite{HGB08}. We describe the potential
scattering by boundary conditions such as one of the components of
the spinor vanishes at a distance $r = R_1$ (of the order of the Bohr radius) 
of the impurity \cite{PGLPN06}. 
A Rashba-like SO interaction exists in the region $R_1 \le r
\le R_2$ (region I), and there is neither potential nor spin orbit
interaction for $r>R_2$, region II ($R_2$ if of the order of the 
carbon-carbon distance).

The wavefunctions in region I can be written as a superposition of
angular harmonics:
%\begin{widetext}
\begin{eqnarray}
\Psi_n ( r , \theta ) &\equiv A_+ \left[ \left(
\begin{array}{c} c_+ J_n ( k_+ r ) e^{i n \theta} \\ i c_- J_{n+1} (
k_+ r ) e^{i (n+1) \theta}
\end{array} \right) \Big| \Big\uparrow \Big\rangle \right.
 + \nonumber \\ &+ \left. \left(
\begin{array}{c} i c_- J_{n+1}
 ( k_+ r ) e^{i (n+1) \theta} \\ - c_+ J_{n+2} ( k_+ r )
e^{i (n+2) \theta} \end{array} \right) \Big| \Big\downarrow \Big\rangle \right] + \nonumber \\
&+ B_+ \left[ \left( \begin{array}{c} c_+ Y_n ( k_+ r ) e^{i n \theta} \\
i c_- Y_{n+1} ( k_+ r ) e^{i (n+1) \theta}
\end{array} \right) \Big| \Big\uparrow \Big\rangle
 + \right. \nonumber \\ &+ \left. \left(
\begin{array}{c} i c_- Y_{n+1}
 ( k_+ r ) e^{i (n+1) \theta} \\ - c_+ Y_{n+2} ( k_+ r )
e^{i (n+2) \theta} \end{array} \right) \Big| \Big\downarrow \Big\rangle \right] + \nonumber \\
&+ A_- \left[ \left( \begin{array}{c} c'_- J_n ( k_- r ) e^{i
\theta}
\\ i c'_+ J_{n+1} ( k_- r ) e^{i (n+1) \theta}
\end{array} \right) \Big| \Big\uparrow \Big\rangle
 - \right. \nonumber \\ &- \left. \left(
\begin{array}{c} i c'_+ J_{n+1}
 ( k_- r ) e^{i (n+1) \theta} \\ - c'_- J_{n+2} ( k_- r )
e^{i (n+2) \theta} \end{array} \right) \Big| \Big\downarrow \Big\rangle \right] + \nonumber \\
&+ B_- \left[ \left( \begin{array}{c} c'_- Y_n ( k_- r ) e^{i n \theta} \\
i c'_+ Y_{n+1} ( k_- r ) e^{i (n+1) \theta}
\end{array} \right) \Big| \Big\uparrow \Big\rangle
 - \right. \nonumber \\ &- \left. \left(
\begin{array}{c} i c'_+ Y_{n+1}
 ( k_- r ) e^{i (n+1) \theta} \\ - c'_- Y_{n+2} ( k_- r )
e^{i (n+2) \theta} \end{array} \right) \Big| \Big\downarrow
\Big\rangle \right]
\end{eqnarray}
%\end{widetext}
where $|\uparrow\rangle$ and $|\downarrow\rangle$ are the spin states.
The functions $J_n ( x ) , Y_n (x)$ are Bessel functions of order
$n$, and:
\begin{eqnarray}
\epsilon &=& \pm \Delta_{so}^{I}/2 + \sqrt{v_F^2 k_\pm^2 +
(\Delta_{so}^{I}/2)^2} 
\label{epi}
\\ 
c_\pm &=& \sqrt{1/2 \pm \Delta_{so}^{I}/(4 \sqrt{v_F^2 k_+^2 +
(\Delta_{so}^{I}/2)^2})} 
\\ 
c'_\pm &=& \sqrt{1/2 \pm
\Delta_{so}^{I}/(4 \sqrt{v_F^2 k_-^2 + (\Delta_{so}^{I}/2)^2})}
\end{eqnarray}
$\epsilon$ is the energy of the scattered electron ($k_{\pm}$ is defined through (\ref{epi})).

The wavefunctions outside the region affected by the impurity,
$r>R_2$, can be written as:
\begin{align}
\Psi_n ( r , \theta ) &\equiv   \left( \begin{array}{c} J_n ( k r )
e^{i n \theta} \\ i  J_{n+1} ( k r ) e^{i (n+1) \theta}
\end{array} \right) \Big| \Big\uparrow \Big\rangle
 + \nonumber \\ &+C_\uparrow \left(
\begin{array}{c}  Y_{n}
 ( k r ) e^{i n \theta} \\ i  Y_{n+1} ( k r )
e^{i (n+1) \theta} \end{array} \right) \Big| \Big\uparrow
\Big\rangle + \nonumber \\ &+ C_\downarrow \left(
\begin{array}{c}  Y_{n+1}
 ( k r ) e^{i (n+1) \theta} \\ i  Y_{n+2} ( k r )
e^{i (n+2) \theta}  \end{array} \right) \Big|  \Big\downarrow \Big\rangle \nonumber \\
\end{align}
and: $\epsilon = v_F k$. 
The boundary conditions at $r=R_1$ and $r=R_2$ lead to the
equations:
\begin{widetext}
\begin{align}
c_+ A_+ J_n ( k_+ R_1 ) + c_+ B_+ Y_n ( k_+ R_1 ) + c'_- A_- J_n (
k_- R_1 ) + c'_-  B_- Y_n ( k_- R_ 1 ) &= 0 \nonumber \\
c_- A_+ J_{n+1} ( k_+ R_1 ) + c_- B_+ Y_{n+1} ( k_+ R_1 ) + c'_+ A_-
J_{n+1} ( k_- R_1 ) + c'_+  B_- Y_{n+1} ( k_- R_1 ) &= 0 \nonumber \\
c_+ A_+ J_n ( k_+ R_2 ) + c_+ B_+ Y_n ( k_+ R_2 ) + c'_- A_- J_n (
k_- R_2 ) + c'_-  B_- Y_n ( k_- R_2 ) &=  J_n ( k R_2 ) + C_\uparrow
Y_n ( k R_2 ) \nonumber \\
c_- A_+ J_{n+1} ( k_+ R_2 ) + c_- B_+ Y_{n+1} ( k_+ R_2 ) + c'_+ A_-
J_{n+1} ( k_- R_2 ) + c'_+  B_- Y_{n+1} ( k_- R_2 ) &=  J_{n+1} ( k
R_2 ) + C_\uparrow Y_{n+1} ( k R_2 ) \nonumber \\
c_- A_+ J_{n+1} ( k_+ R_2 ) + c_- B_+ Y_{n+1} ( k_+ R_2 ) - c'_+ A_-
J_{n+1} ( k_- R_2 ) - c'_+  B_- Y_{n+1} ( k_- R_2 ) &=
C_\downarrow
Y_{n+1} ( k R_2 ) \nonumber \\
c_+ A_+ J_{n+2} ( k_+ R_2 ) + c_+ B_+ Y_{n+2} ( k_+ R_2 ) - c'_- A_-
J_{n+2} ( k_- R_2 ) - c'_-  B_- Y_{n+2} ( k_- R_2 ) &= C_\downarrow
Y_{n+2} ( k R_2 )
\end{align}
\end{widetext}
These six equations allow us to obtain the coefficients $A_\pm ,
B_\pm , C_\uparrow$ and $C_\downarrow$. In the absence of the
SO interaction, we have $A_+ = A_- , B_+ = B_- ,
C_\downarrow =0$ and $C_\uparrow = - J_n ( k R_1 ) / Y_n ( k R_1 )$.

\begin{figure}
\begin{center}
\includegraphics[width=8cm,angle=0]{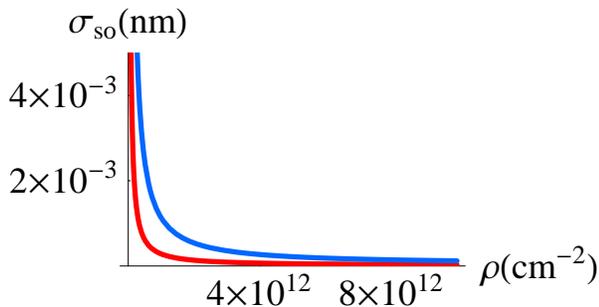}
\caption[fig]{(Color online). Cross section for a spin flip process
for a defect as described in the text. The parameters used are $R_1
= 1$\AA \, $R_2 = 2$\AA \, and $\Delta_{so}^I=1$meV (blue) and
$\Delta_{so}^I = 2$meV (red) .} \label{cross_section}
\end{center}
\end{figure}

We show in Fig.~\ref{cross_section} the results for the cross
section for spin flip processes, determined by $|C_{\downarrow}|^2 / k_F$. 
The main contribution arises from the $n=0$ channel. For
comparison, the elastic cross section, calculated in the same way,
is $\sigma_{el} \approx k_F^{-1}$. This is about three of magnitude
larger than the spin-flip cross section due to the spin orbit
coupling. Hence, the spin relaxation length is $10^3$ times the
elastic mean free path \cite{HGB08}. We obtain a mean free path of
about 1 $\mu m$, in reasonable agreement with the experimental
results in ref.~\cite{HJPJW07t}. This value depends quadratically on
$\Delta_{so}^I(A)$. For a finite, but small, concentration of impurities,
our results scale with the impurity concentration and hence the spin
flip processes should increase roughly linearly with impurity coverage
in transport experiments in systems like graphane \cite{graphane}.

In summary, we have shown that the impurity induced, lattice driven, 
SO coupling in graphene can be of the order of the atomic spin orbit
coupling and comparable to what is found in diamond and zinc-blend
semiconductors. The value of the
SO coupling depends on how much the carbon atom which is hybridized
with the impurity displaces from the plane inducing a sp$^3$ 
hybridization. We have calculated the spin-flip cross section due
to SO coupling for the impurity and shown that it agrees with 
recent experiments. This results indicates that there are substantial
amounts of hybridized impurities in graphene, even under ultra-clean 
high vacuum conditions. Experiments where the impurity coverage is well
controlled can provide a ``smoking-gun'' test of our predictions.

We thank illuminating discussions with D. Huertas-Hernando and A. Brataas. 
AHCN acknowledges the partial support of the U.S. Department of Energy under grant DE-FG02-08ER46512. FG acknowledges support from MEC (Spain) 
through grant FIS2005-05478-C02-01 and CONSOLIDER CSD2007-00010, by
the Comunidad de Madrid, through CITECNOMIK, CM2006-S-0505-ESP-0337.

\bibliography{spin_orbit_nanotubes_5}

\begin{thebibliography}{24}
\expandafter\ifx\csname natexlab\endcsname\relax\def\natexlab#1{#1}\fi
\expandafter\ifx\csname bibnamefont\endcsname\relax
  \def\bibnamefont#1{#1}\fi
\expandafter\ifx\csname bibfnamefont\endcsname\relax
  \def\bibfnamefont#1{#1}\fi
\expandafter\ifx\csname citenamefont\endcsname\relax
  \def\citenamefont#1{#1}\fi
\expandafter\ifx\csname url\endcsname\relax
  \def\url#1{\texttt{#1}}\fi
\expandafter\ifx\csname urlprefix\endcsname\relax\def\urlprefix{URL }\fi
\providecommand{\bibinfo}[2]{#2}
\providecommand{\eprint}[2][]{\url{#2}}

\bibitem[{\citenamefont{{Novoselov {\it et al.}}}(2004)}]{Netal04t}
\bibinfo{author}{\bibfnamefont{K.~S.} \bibnamefont{{Novoselov {\it et al.}}}},
  \bibinfo{journal}{Science} \textbf{\bibinfo{volume}{306}},
  \bibinfo{pages}{666} (\bibinfo{year}{2004}).

\bibitem[{\citenamefont{Geim and Novoselov}(2007)}]{geim_review}
\bibinfo{author}{\bibfnamefont{A.~K.} \bibnamefont{Geim}} \bibnamefont{and}
  \bibinfo{author}{\bibfnamefont{K.~S.} \bibnamefont{Novoselov}},
  \bibinfo{journal}{Nature Materials} \textbf{\bibinfo{volume}{6}},
  \bibinfo{pages}{183} (\bibinfo{year}{2007}).

\bibitem[{\citenamefont{{{Castro Neto} {\it et al.}}}(2009)}]{NGPNG08}
\bibinfo{author}{\bibfnamefont{A.~H.} \bibnamefont{{{Castro Neto} {\it et
  al.}}}}, \bibinfo{journal}{Rev. Mod. Phys.} \textbf{\bibinfo{volume}{81}},
  \bibinfo{pages}{109} (\bibinfo{year}{2009}).

\bibitem[{\citenamefont{{Chen {\it et al.}}}(2008)}]{CJFWI08}
\bibinfo{author}{\bibfnamefont{J.~H.} \bibnamefont{{Chen {\it et al.}}}},
  \bibinfo{journal}{Nat. Phys.} \textbf{\bibinfo{volume}{4}},
  \bibinfo{pages}{377} (\bibinfo{year}{2008}).

\bibitem[{\citenamefont{{Blake {\it et al.}}}(2008)}]{blake08}
\bibinfo{author}{\bibfnamefont{P.}~\bibnamefont{{Blake {\it et al.}}}}
  (\bibinfo{year}{2008}), \eprint{arXiv:0810.4706}.

\bibitem[{\citenamefont{{Elias {\it et al.}}}(2009)}]{graphane}
\bibinfo{author}{\bibfnamefont{D.~C.} \bibnamefont{{Elias {\it et al.}}}},
  \bibinfo{journal}{Science} \textbf{\bibinfo{volume}{323}},
  \bibinfo{pages}{610} (\bibinfo{year}{2009}).

\bibitem[{\citenamefont{{Tombros {\it et al.}}}(2007)}]{HJPJW07t}
\bibinfo{author}{\bibfnamefont{N.}~\bibnamefont{{Tombros {\it et al.}}}},
  \bibinfo{journal}{Nature} \textbf{\bibinfo{volume}{448}},
  \bibinfo{pages}{571} (\bibinfo{year}{2007}).

\bibitem[{\citenamefont{{Tombros {\it et al.}}}(2008)}]{TTVJJvW08t}
\bibinfo{author}{\bibfnamefont{N.}~\bibnamefont{{Tombros {\it et al.}}}},
  \bibinfo{journal}{Phys. Rev. Lett.} \textbf{\bibinfo{volume}{101}},
  \bibinfo{pages}{046601} (\bibinfo{year}{2008}).

\bibitem[{\citenamefont{Huertas-Hernando
  et~al.}(2008)\citenamefont{Huertas-Hernando, Guinea, and Brataas}}]{HGB08}
\bibinfo{author}{\bibfnamefont{D.}~\bibnamefont{Huertas-Hernando}},
  \bibinfo{author}{\bibfnamefont{F.}~\bibnamefont{Guinea}}, \bibnamefont{and}
  \bibinfo{author}{\bibfnamefont{A.}~\bibnamefont{Brataas}}
  (\bibinfo{year}{2008}), \eprint{arXiv:0812.1921}.

\bibitem[{\citenamefont{Huertas-Hernando
  et~al.}(2006)\citenamefont{Huertas-Hernando, Guinea, and Brataas}}]{HGB06}
\bibinfo{author}{\bibfnamefont{D.}~\bibnamefont{Huertas-Hernando}},
  \bibinfo{author}{\bibfnamefont{F.}~\bibnamefont{Guinea}}, \bibnamefont{and}
  \bibinfo{author}{\bibfnamefont{A.}~\bibnamefont{Brataas}},
  \bibinfo{journal}{Phys. Rev. B} \textbf{\bibinfo{volume}{74}},
  \bibinfo{pages}{155426} (\bibinfo{year}{2006}).

\bibitem[{\citenamefont{Kane and Mele}(2005)}]{KM05}
\bibinfo{author}{\bibfnamefont{C.~L.} \bibnamefont{Kane}} \bibnamefont{and}
  \bibinfo{author}{\bibfnamefont{E.~J.} \bibnamefont{Mele}},
  \bibinfo{journal}{Phys. Rev. Lett.} \textbf{\bibinfo{volume}{95}},
  \bibinfo{pages}{226801} (\bibinfo{year}{2005}).

\bibitem[{\citenamefont{Kane and Mele}(2006)}]{KM07}
\bibinfo{author}{\bibfnamefont{C.}~\bibnamefont{Kane}} \bibnamefont{and}
  \bibinfo{author}{\bibfnamefont{E.}~\bibnamefont{Mele}},
  \bibinfo{journal}{Science} \textbf{\bibinfo{volume}{314}},
  \bibinfo{pages}{1692} (\bibinfo{year}{2006}).

\bibitem[{\citenamefont{Duplock et~al.}(2004)\citenamefont{Duplock, Scheffler,
  and Lindan}}]{DSL07}
\bibinfo{author}{\bibfnamefont{E.~J.} \bibnamefont{Duplock}},
  \bibinfo{author}{\bibfnamefont{M.}~\bibnamefont{Scheffler}},
  \bibnamefont{and} \bibinfo{author}{\bibfnamefont{P.~J.}
  \bibnamefont{Lindan}}, \bibinfo{journal}{Phys. Rev. Lett.}
  \textbf{\bibinfo{volume}{92}}, \bibinfo{pages}{225502}
  (\bibinfo{year}{2004}).

\bibitem[{\citenamefont{Yu and Cardona}(2005)}]{cardona}
\bibinfo{author}{\bibfnamefont{P.~Y.} \bibnamefont{Yu}} \bibnamefont{and}
  \bibinfo{author}{\bibfnamefont{M.}~\bibnamefont{Cardona}},
  \emph{\bibinfo{title}{Fundamentals of Semiconductors: Physics and Materials
  Properties}} (\bibinfo{publisher}{Springer, New York}, \bibinfo{year}{2005}).

\bibitem[{\citenamefont{Serrano et~al.}(2000)\citenamefont{Serrano, Cardona,
  and Ruf}}]{SCR00}
\bibinfo{author}{\bibfnamefont{J.}~\bibnamefont{Serrano}},
  \bibinfo{author}{\bibfnamefont{M.}~\bibnamefont{Cardona}}, \bibnamefont{and}
  \bibinfo{author}{\bibfnamefont{T.}~\bibnamefont{Ruf}},
  \bibinfo{journal}{Solid St. Commun.} \textbf{\bibinfo{volume}{113}},
  \bibinfo{pages}{411} (\bibinfo{year}{2000}).

\bibitem[{\citenamefont{Elliot}(1954)}]{Elliot54}
\bibinfo{author}{\bibfnamefont{P.~G.} \bibnamefont{Elliot}},
  \bibinfo{journal}{Phys. Rev.} \textbf{\bibinfo{volume}{96}},
  \bibinfo{pages}{266} (\bibinfo{year}{1954}).

\bibitem[{\citenamefont{Yafet}(1963)}]{Yafet63}
\bibinfo{author}{\bibfnamefont{Y.}~\bibnamefont{Yafet}}, in
  \emph{\bibinfo{booktitle}{Solid State Physics, vol 13}}, edited by
  \bibinfo{editor}{\bibnamefont{ed.~by F.~Seitz}} \bibnamefont{and}
  \bibinfo{editor}{\bibfnamefont{D.}~\bibnamefont{Turnbull}}
  (\bibinfo{publisher}{Academic, New York}, \bibinfo{year}{1963}).

\bibitem[{\citenamefont{Harrison}(1980)}]{harrison}
\bibinfo{author}{\bibfnamefont{W.~A.} \bibnamefont{Harrison}},
  \emph{\bibinfo{title}{Solid State Theory}} (\bibinfo{publisher}{Dover, New
  York}, \bibinfo{year}{1980}).

\bibitem[{\citenamefont{Ostrovsky et~al.}(2006)\citenamefont{Ostrovsky, Gornyi,
  and Mirlin}}]{OGM06}
\bibinfo{author}{\bibfnamefont{P.~M.} \bibnamefont{Ostrovsky}},
  \bibinfo{author}{\bibfnamefont{I.~V.} \bibnamefont{Gornyi}},
  \bibnamefont{and} \bibinfo{author}{\bibfnamefont{A.~D.}
  \bibnamefont{Mirlin}}, \bibinfo{journal}{Phys. Rev. B}
  \textbf{\bibinfo{volume}{74}}, \bibinfo{pages}{235443}
  (\bibinfo{year}{2006}).

\bibitem[{\citenamefont{Hentschel and Guinea}(2007)}]{HG07}
\bibinfo{author}{\bibfnamefont{M.}~\bibnamefont{Hentschel}} \bibnamefont{and}
  \bibinfo{author}{\bibfnamefont{F.}~\bibnamefont{Guinea}},
  \bibinfo{journal}{Phys. Rev. B} \textbf{\bibinfo{volume}{76}},
  \bibinfo{pages}{115407} (\bibinfo{year}{2007}).

\bibitem[{\citenamefont{Novikov}(2007)}]{N07}
\bibinfo{author}{\bibfnamefont{D.~S.} \bibnamefont{Novikov}},
  \bibinfo{journal}{Phys. Rev. B} \textbf{\bibinfo{volume}{76}},
  \bibinfo{pages}{245435} (\bibinfo{year}{2007}).

\bibitem[{\citenamefont{Katsnelson and Novoselov}(2007)}]{KN07}
\bibinfo{author}{\bibfnamefont{M.~I.} \bibnamefont{Katsnelson}}
  \bibnamefont{and} \bibinfo{author}{\bibfnamefont{K.~S.}
  \bibnamefont{Novoselov}}, \bibinfo{journal}{Solid State Commun.}
  \textbf{\bibinfo{volume}{143}}, \bibinfo{pages}{3} (\bibinfo{year}{2007}).

\bibitem[{\citenamefont{Guinea}(2008)}]{G08}
\bibinfo{author}{\bibfnamefont{F.}~\bibnamefont{Guinea}},
  \bibinfo{journal}{Journ. Low Temp. Phys.} \textbf{\bibinfo{volume}{153}},
  \bibinfo{pages}{359} (\bibinfo{year}{2008}).

\bibitem[{\citenamefont{{Pereira {\it et al.}}}(2006)}]{PGLPN06}
\bibinfo{author}{\bibfnamefont{V.~M.} \bibnamefont{{Pereira {\it et al.}}}},
  \bibinfo{journal}{Phys. Rev. Lett.} \textbf{\bibinfo{volume}{96}},
  \bibinfo{pages}{036801} (\bibinfo{year}{2006}).

\end{thebibliography}
\end{document}